\documentstyle[12pt]{article}

\advance\textwidth 2.15cm
\advance\oddsidemargin -1cm
\advance\textheight 1.5cm
\advance\topmargin -1cm

\newcommand\beq{\begin{equation}}
\newcommand\eeq{\end{equation}}
\def\beqa{\begin{eqnarray}}
\def\eeqa{\end{eqnarray}}
\def\bega{\begin{array}}
\def\enda{\end{array}}
\def\be{\[}
\def\ee{\]}
\def\non{{\nonumber}}

\def\sp{\hspace{.8cm}}

\def\bq{{\bar q}}
\def\vb{\vphantom{b}}

\def\l{{\langle}}
\def\r{{\rangle}}
\def\dag#1{{\dagger_#1}}
\def\H#1{H^{(#1)}}
\def\sHa{{\cal H}}
\def\sH#1{{\cal H}^{(#1)}}

\def\c#1{_{,#1} }  

\def\ph#1{\phi^{(#1)}}
\def\ps#1{\psi^{(#1)}}

\def\arcsinh{{\rm arcsinh}}
\def\ker{{\rm ker}}

\def\lt{{\tt <}}
\def\rt{{\tt >}}

\def\D{{\cal D}}

\def\M{{\cal M}}
\def\Q{{\cal Q}}
\def\U{{\cal U}}

\def\a{\alpha}
\def\b{\beta}
\def\g{\gamma}
\def\ga{\gamma}
\def\eps{\epsilon}
\def\la{\lambda}
\def\w{\omega}

\def\d{\partial}

\begin{document}

\title{Canonical Transformations in Quantum Mechanics}
\author{Arlen Anderson\thanks{arley@ic.ac.uk}\\
Department of Physics\\
McGill University\\
Ernest Rutherford Building\\
Montr\'eal PQ Canada H3A 2T8\\
and\\
Blackett Laboratory\thanks{Present address.}\\
Imperial College\\
Prince Consort Road\\
London SW7 2BZ UK\\
and\\
Institute of Theoretical Physics\\
Univ. of California\\
Santa Barbara CA 93106 }
\date{May 20, 1992\\ Revised: May 10, 1993}
\maketitle

\vspace{-15.75cm}
\hfill McGill 92-29

\hfill Imperial-TP-92-93-31

\hfill NSF-ITP-93-61

\hfill hep-th/9305054
\vspace{13cm}

\begin{abstract}
Quantum canonical transformations are defined algebraically outside of a
Hilbert space context. This generalizes the quantum canonical
transformations of Weyl and Dirac to include non-unitary transformations.
The importance of non-unitary transformations for constructing solutions of
the Schr\"odinger equation is discussed. Three elementary canonical
transformations are shown both to have quantum implementations as finite
transformations and to generate, classically and infinitesimally, the full
canonical algebra. A general canonical transformation can be realized
quantum mechanically as a product of these transformations. Each
transformation corresponds to a familiar tool used in solving differential
equations, and the procedure of solving a differential equation is
systematized by the use of the canonical transformations. Several examples
are done to illustrate the use of the canonical transformations.
\end{abstract}
PACS: 03.65.Ca, 03.65.Ge
\newpage

\section{Introduction}

Canonical transformations are a powerful tool of classical mechanics whose
strength has not been fully realized in quantum mechanics. Canonical
transformations are already widely used, at least implicitly, because as
Dirac\cite{Dir} and Weyl\cite{Wey} emphasized the unitary transformations
are canonical. Aside from the linear canonical transformations, which have
been well understood for many years\cite{lin}, comparatively little work
(but see, e.g., \cite{Mos,LeS,Dee}) has been done directly on quantum
canonical transformations as nonlinear changes of the non-commuting phase
space variables. Progress has been inhibited because of the mistaken belief
that quantum canonical transformations must be unitary.

Classically, a canonical transformation is a change of the phase space
variables $(q_c,p_c)\mapsto (q'_c(q_c,p_c),p'_c(q_c,p_c))$ which preserves
the Poisson bracket $\{q_c,p_c\}=1=\{q'_c,p'_c\}$. Born, Heisenberg, and
Jordan\cite{BHJ} propose as the natural definition of
a quantum canonical transformation: a change of the non-commuting phase
space variables
\beq
\label{def}
q\mapsto q'(q,p),\sp p\mapsto p'(q,p),
\eeq
which preserves the Dirac bracket (canonical commutation relations)
\beq
[q,p]=i=[q'(q,p),p'(q,p)].
\eeq
Such a transformation is implemented by a function $C(q,p)$ such that
\beq
q'(q,p)= C q C^{-1},\sp p'(q,p)=C p C^{-1}.
\eeq
This definition applies at the purely algebraic level, and no Hilbert
space or inner product need be mentioned.  As a consequence, the
transformation is itself neither unitary nor non-unitary.  The immediate
implication is that quantum canonical transformations may be non-unitary,
a generalization beyond Dirac's and Weyl's consideration.

Canonical transformations serve three primary purposes: for evolution, to
prove physical equivalence, and to solve a theory. Classically, these blur
together. Quantum mechanically, however, they are distinct. Evolution is
produced by unitary transformations\cite{Dir,Wey}. Physical equivalence is
proven with isometric transformations, which extend the definition of
unitary transformations to norm-preserving isomorphisms between different
Hilbert spaces\cite{And1}. (It is useful to distinguish unitary and
isometric because for many physicists the working definition of a unitary
transformation $U$ is $U^\dagger U=1$, and this is not true of isometric
transformations.) Solution of a theory is accomplished with general
canonical transformations and may involve non-unitary
transformations\cite{And2}.

Mello and Moshinsky\cite{Mos} raise three issues concerning the definition
(\ref{def}) of a quantum canonical transformation. First, the ordering of
$q'(q,p)$ and $p'(q,p)$ must be given, so that they are well-defined.
Second, when the transformation is represented by operators, inverse and
fractional powers of differential operators may appear, and these must be
defined. Third, the transformation may be non-unitary, and the sense of
this must be understood. This paper addresses these concerns.

A quantum phase space is introduced which consists of pairs of canonically
conjugate elements of a non-commutative algebra $\U$ constructed from the
phase space variables $q$, $p$. Objects like the Hamiltonian which are
ordered combinations of $q$, $p$ are elements of $\U$. Each element of $\U$
defines a canonical transformation, so $\U$ may be identified as the
quantum canonical group. It is assumed to be a topological transformation
group which acts transitively on itself by the adjoint action. Expressions
like $p^{-\a}$ ($\a$ an arbitrary complex number) are well-defined as
elements of $\U$. Because the canonical commutation relations are imposed
as relations on $\U$, every function in the algebra has a well-defined
ordering. The quantum phase space is thus defined algebraically without
specifying an inner product or Hilbert space structure---canonical
transformations made upon it preserve the quantum structure of the
canonical commutation relations, but these transformations are neither
unitary nor non-unitary.

The phase space variables $(q,p)$ are represented as operators
$(\check{q},\check{p})\equiv (q,-i\d_q)$ which act on functions $\psi(q)$
on configuration space, again without specifying the Hilbert space
structure. Inverse and fractional powers are understood in a sense
analogous to that of pseudo-differential operators\cite{pseudo}. The
ordering implicit in elements of $\U$ gives a well-defined ordering to the
corresponding quantum operators. Since the operators are defined outside of
a Hilbert space, they define a transformation on all solutions of the
Schr\"odinger equation, including non-normalizable ones.

Once the Hilbert space of physical states is specified, one may find that
the kernel of a particular canonical transformation lies in the
Hilbert space or that the normalization of states changes under the
transformation.  In these cases, the canonical transformation is
non-unitary, but nevertheless it may have proven useful in constructing
the explicit representation of solutions of the Schr\"odinger equation.

Classically, a theory is solved with canonical transformations by
transforming the Hamiltonian to a simpler one whose
equations of motion can be solved. The implementation of this same program
quantum mechanically is the most significant application of the
approach to quantum canonical transformations described here and
will be the focus.  It will be shown that a general
quantum canonical transformation can be constructed as a product of
elementary canonical transformations of known behavior, as conjectured by
Levyraz and Seligman\cite{LeS}.  These elementary canonical
transformations each correspond to a familiar tool used in solving
differential equations:  change of variables, extracting a function of
the independent variables from the dependent variable, and Fourier
transform.  The procedure of solving a differential equation is
systematized by the use of the elementary canonical transformations.
More sophisticated tools for solving differential equations, including raising
and lowering operators\cite{InH}, supersymmetry\cite{susy},
intertwining operators\cite{And3,And4}, and
Lie algebraic methods\cite{lam}, may also be shown to be canonical
transformations.

The quantum canonical transformations provide a unified approach to the
integrability of quantum systems.  One may define as quantum integrable
(``in the sense of canonical transformations'') those problems whose
general solution can be constructed as a finite product of
elementary canonical transformations.
In view of the fact
that the standard tools for integrating differential equations are among
the canonical transformations, most if not all of the known soluble
equations are in this class.

The unifying nature of the canonical transformations is a
consequence of the size of the canonical group.  The canonical group
provides the maximum freedom to transform the form of a differential equation
while preserving the relation between the derivative and the coordinate.
The approach of Lie to the solution of differential equations involves
constructing symmetries that transform an equation into itself.  These
finite dimensional Lie groups are subgroups of the canonical group.
Generalized symmetries, which preserve the broad form of an equation, say
that it remain a second order derivative plus a potential (such as
appears in the Lax formulation of the Korteweg-deVries equation), give
rise to infinite-dimensional symmetry groups whose algebras are Kac-Moody
algebras.  These are subgroups of the canonical group as well.

The outline of the paper is as follows. The treatment begins with the
definition of the quantum phase
space and the algebra $\U$ (Section~\ref{sqpsp}).  An effort is made to
be precise about the definition of $\U$ as this determines the class of
functions that may be used for making canonical transformations.
Next, the elements of $\U$ are represented as operators acting on
functions on configuration space (Section~\ref{srep}). The definition
of a quantum canonical transformation is then reviewed, and some basic
properties established (Section~\ref{sqct}).  The conditions under which
a canonical transformation defines a physical equivalence are found
(Section~\ref{unieq}).

This is followed by a
discussion of classical infinitesimal canonical transformations. Three
elementary canonical transformations are
introduced, and two additional composite elementary
transformations are constructed. It is shown that, classically,
the infinitesimal forms of these transformations generate the full algebra
of canonical transformations. The inference is drawn that in principle by
using their quantum implementations, any quantum canonical transformation
can be made (Section~\ref{inftr}). Next, the quantum implementation of
the elementary canonical transformations is given.  An example
illustrating their use is given by computing the exponential of the
infinitesimal generator of point canonical transformations
(Section~\ref{qimpl}).

The next section is devoted to examples.  First, the linear canonical
transformations are constructed as products of elementary
transformations (Section~\ref{linct}).  The free particle and the
harmonic oscillator are then done to illustrate various aspects
of the use of canonical transformations (Sections~\ref{sfree}-\ref{shoprop}).
Of note is the
solution of the time-dependent harmonic oscillator by an approach which
suggests that many time-independent integrable potentials can be
generalized to contain time-dependent parameters without sacrificing
integrability (Section~\ref{stdho}).  The first order intertwining operator
is shown to be a canonical transformation; this allows the construction
of the recursion relations and Rodrigues' formulae for all equations that
are essentially hypergeometric (Section~\ref{sio}).  An example involving
a form of Bessel's equation is done to emphasize the difference between
the classical and quantum canonical transformations which simplify a
Hamiltonian to a given one (Section~\ref{clvq}).  Finally, the problem of
a particle propagating on a higher-dimensional sphere is solved as an
illustration of the use of intertwining operators as canonical
transformations (Section~\ref{nsph}).

Readers primarily interested in applications may focus on the sections
giving the definition of quantum canonical transformations (Sec.~\ref{sqct})
and the implementation of the elementary canonical transformations
(Sec.~\ref{qimpl}) before moving to the examples of Sec.~\ref{Examples}.

\section{Formal aspects}

\subsection{The Quantum Phase Space}
\label{sqpsp}

Before discussing quantum canonical transformations, it is important to
define the space on which they will act, that is, the quantum phase space.
To set the stage, consider the classical
phase space for an unconstrained system of $n$ degrees of freedom.
Classically, the phase space is the vector space $R^{2n}$ with coordinates
$(q_1,\ldots,q_n,p_1,\ldots,p_n)$. The phase space can be extended to
include time $q_0$ and its conjugate momentum $p_0$\cite{Lan}. The Hamiltonian
$H(q_1,\ldots,q_n,p_1,\ldots,p_n,q_0)$ is a function on the extended phase
space, as are all possible observables, such as the position of the center
of mass, the total angular momentum, the momentum of the $k^{\rm th}$
degree of freedom, etc. On the phase space, there is a symplectic form
$\sum_{k=0}^n dq_k\wedge dp_k$ which is reflected in the Poisson bracket of
functions on phase space
$$\{f,g\}= \sum_{k=0}^n \d_{q_k} f \d_{p_k} g
-\d_{p_k} f \d_{q_k} g .$$
A canonical transformation is a change of
coordinates on extended phase space which preserves the symplectic form, or
equivalently, the Poisson bracket relations.

Quantum mechanically, there is a parallel structure but with several
crucial differences. The phase space variables $\{q_k \}$, $\{p_k \}$
($k=0,\ldots,n$) are
members of a non-commutative algebra $\U$. As such, they cannot be thought
of as coordinates on a vector space in the usual sense. Instead the
parallel to the classical case is drawn one level higher between classical
observables and elements of $\U$. Broadly speaking, the elements of $\U$
are sums of ordered products of the $\{q_k \}$, $\{p_k \}$ (and their
algebraic inverses $\{q_k^{-1} \}$, $\{p_k^{-1} \}$) which may be
expressed $f\lt q_0,\ldots,q_n,p_0,\ldots,p_n \rt$. Two expressions related
by repeated application of the canonical commutation relations are
equivalent and represent the same element in $\U$. Objects like the quantum
Hamiltonian are the analog of the classical observables. The $\{q_k \}$,
$\{p_k \}$ act in a sense as coordinates as they label the elements of $\U$
with respect to a set of preferred elements, namely themselves. The quantum
phase space $\M$ is defined to be the set of canonically conjugate quantum
variables
\beq
\M=\{ (\{x_j\},\{ p_{x_k}\} )\in \bigoplus_1^{2n+2} \U  \,|\,
[x_j,p_{x_k}]=i\delta_{jk},
\ [x_j,x_k]=0,\  [p_{x_j},p_{x_k}]=0 \}.
\eeq
The specification of the commutation relations is analogous to defining
the Poisson bracket structure or symplectic form.
A canonical transformation is a mapping $C\in\U$ from $\M$ to $\M$ given by
\beq
C: (\{x_j\},\{ p_{x_k}\} ) \mapsto (\{C x_j C^{-1}\},\{ Cp_{x_k} C^{-1}\} ).
\eeq
By construction, the mapping $C$ preserves the canonical commutation
relations.

Classically, any function of the classical variables
$$F(q_1,\ldots,q_n,p_1,\ldots,p_n,q_0,p_0)$$
is an observable. Precisely what
class of functions is allowed is vague, but intuitively one wants smooth
($C^\infty$) functions on $R^{2n+2}$, with possibly a countable set of
singularities.  The task at hand is to characterize the elements of
$\U$.

First a point of terminology: quantum mechanically, the word ``observable''
cannot be used to refer to an element of $\U$ because it is already
defined in an incompatible way. The word ``function'' is also inappropriate
because strictly speaking a function is a mapping from a domain to a range,
and the elements of $\U$ do not take values in some domain and are not
themselves mappings. The natural choice would be the term ``symbol,''
denoting an element of a ring, but this word is already in common usage in
a related but distinct way for pseudo-differential operators\cite{pseudo}.

With reluctance, the new term ``notion'' is proposed. The generators $q_k$,
$p_k$ ($k=0,\ldots, n$) are called ``notes,'' marking certain elements of
$\U$ as significant. An element $f\in\U$ having an expression $f\lt
q_0,\ldots,q_n,p_0,\ldots,p_n \rt$ as an ordered combination of
notes is a ``notion.'' The angle brackets are used to emphasize that $f$ is
not a function of the notes, but an ordered expression in terms of them.
Angle brackets and this terminology will only be used in this section. In
later sections, the convenient fiction that elements of $\U$ like the
Hamiltonian are functions of the phase space variables will be maintained
in conformance with common practice. When a notion is represented as
an operator, it is called a ``notable''---observables are a subset of
the notables.

For convenience, let $(q,p)$ denote the phase space variables.  The
extension to the many-variable case is straightforward.

Begin with an intuitive description of $\U$.  First, it must
contain all possible Hamiltonians, as well as all the notions corresponding
to observables.  In particular, since every classical potential is a possible
analog of a quantum potential, $\U$ must contain the notion expressions for
the $C^\infty$ functions of $q$, with singularities, that are allowed
classically.  Classically, this space of functions is topologized by
defining a family of seminorms on open sets of the
domain of the functions\cite{Rud}.  Quantum mechanically, there is no
underlying
domain from which to induce the topology.  One must work directly with
the abstract algebraic expression---the notion---corresponding to a function.
Nevertheless one expects that a topology persists, even if it is
difficult to define it precisely. The existence of this
topology will be assumed.

Secondly, $\U$ must be closed under canonical transformations.  This means
that if $x,C\in\U$ then $CxC^{-1}\in\U$. The stronger assumption is
made that the adjoint action of $C$ is transitive, that is, for every
$x,y\in\U$, there exists a $C\in\U$ such that $CxC^{-1}=y$.  This
property is important for defining functions on $\U$\cite{Dir2}.

Define a notion of $q$
to be those $x\in\U$ such that $[x,q]=0$, denoted $x\lt q \rt$.  There is
a subgroup $\D\subset \U$ which transforms a notion of $q$ to a notion of
$q$.  The notions $x\lt q \rt$ form a normal subgroup
$\Q$ of $\D$. The quotient group $\D/\Q$
is the abstract analog of
the diffeomorphism group.  Part of the difficulty in topologizing $\U$ can be
traced to the fact that it contains the diffeomorphism group as a subgroup,
and the topology of the diffeomorphism group is a deep subject in
itself\cite{Mil}.

{}From this discussion, it is clear that the algebra $\U$ is assumed to be
a topological transformation group\cite{MoZ} which acts transitively on
itself by the adjoint action.
A more precise definition of $\U$ may be given by describing a
construction of it.  One begins with the Weyl algebra\cite{ring} of
polynomials in the notes $q,p$,
with the relation
\beq
q p -p q =i.
\eeq
One adjoins the elements $q^{-1},p^{-1}$, together with the relations
\beq
q^{-1} q= q q^{-1} =1, \sp p^{-1} p= p p^{-1} =1.
\eeq
Additional relations like
\be
q^{-1} p -p q^{-1} = -i q^{-2}
\ee
follow from those given already, as, for example, this one is
$-q^{-1}[q,p] q^{-1}$.

The core $\U_c$ of the algebra $\U$ is given by the ring over the complex
field of formal
Laurent expansions in $q,p$ having only a finite number of non-vanishing
coefficients of negative powers.
A general element $f\in \U_c$ can be
expressed as the notion
\beq
\label{laur}
f\lt q,p \rt = \sum_{ \{n_j,m_j \} } a_{\{n_j,m_j\} } q^{n_1} p^{m_1}
    q^{n_2} p^{n_2} \cdots.
\eeq
These notions could be topologized in terms of open sets in a space
coordinatized by the expansion coefficients.

Using the relations, the notion (\ref{laur}) can be
reordered to put all of the $p$'s on the right.  Representing the $p$'s
as differential operators, the symbol $f(q,\xi)$ of the
corresponding pseudodifferential operator is obtained by taking the
Fourier
transform\cite{pseudo}.  One can go on to prove that the pseudodifferential
operators form an algebra, and the class of symbols has a topology
as $C^\infty$ functions.  This procedure is not followed here because it
is useful not to insist on reordering a notion to obtain the symbol, and
indeed for many notions that will be used, it is not obvious what their
reordering is nor what growth properties their symbols would have.

The algebra $\U$ is completed by closing $\U_c$ under an action of
functional composition.
The space
of canonically conjugate variables is the set
\beq
\M=\{ (x,p_x)\in \U\oplus \U \,|\, x p_x- p_x x=i \}.
\eeq
Each notion $f\lt q,p \rt$ induces a function $\tilde f$ from $\M$ to
$\U$ by
\beq
\label{indef}
\tilde f(x,p_x)\lt q,p \rt=f\lt x\lt q,p\rt ,p_x\lt q,p\rt \rt.
\eeq
As an example, the induced action of the notion $f\lt q,p \rt= q p^3$
on a pair $(x,p_x)$ is
\beq
\tilde f(x,p_x)= x p_x^3.
\eeq
When this action is evaluated on the pair
\be
(x\lt q,p\rt ,p_x\lt q,p\rt)= (q^2, (2q)^{-1} p),
\ee
one has
\beq
\tilde f(x,p_x)\lt q,p \rt= q^2 \left( {1\over 2q} p \right)^3.
\eeq

By the transitivity of the adjoint action of $\U$ on itself, this
function has the important property
\beq
\tilde f(C x C^{-1}, C p_x C^{-1}) = C \tilde f(x,p_x) C^{-1},
\eeq
which defines the value of the function everywhere in $\M$ in terms of
its value at $(q,p)$.  Dirac grappled with the issue of defining a
function of non-commuting variables in the early days of quantum
mechanics\cite{Dir2}, and this was one of his requirements. From
(\ref{indef}), one sees that the
function $C\tilde f(x,p_x) C^{-1}$ is induced by the notion $Cf C^{-1}$.

On the subgroup $\Q$ of notions of $q$, there is a similar induced mapping
from $\U$ to $\U$
\beq
\tilde f(x)\lt q,p \rt= f\lt x\lt q,p \rt \rt.
\eeq
Clearly, closure above induces closure on this subgroup.  Focussing
on this subgroup, one can see the nature of the extension of $\U_c$
produced by the repeated action of the induced
mapping of $\M$ on $\U$.

The requirement of closure introduces three new types of elements.
First, the induced action of $f=q^{-1}$ on $x\in\U$ gives the algebraic
inverse elements $x^{-1}$.
Acting on $q$, one has $\tilde f(q)=q^{-1}$.  Given $C$ such
that $x=C q C^{-1}$, one has
\beq
Cq^{-1} C^{-1}= C\tilde f(q) C^{-1} = \tilde f(x)\equiv x^{-1}.
\eeq
But applying $C$ to
$q^{-1} q=q q^{-1}=1$ gives
\beq
C q^{-1} C^{-1} x= x C q^{-1} C^{-1} =1.
\eeq
Therefore $x^{-1}\equiv Cq^{-1} C^{-1}$ is the algebraic inverse of $x$.
Uniqueness of the inverse in $\U$ then
establishes correspondences like
\be
(1-q)^{-1}=\sum_{n=0}^\infty q^n.
\ee

Secondly, the induced action of a formal Laurent expansion having an
infinite number of non-vanishing coefficients of positive powers on
$q^{-1}$ gives a formal expansion having an infinite number of
non-vanishing coefficients of negative powers.  For example, the induced action
of $\exp(-q^2)\equiv \sum_{n=0}^{\infty} (-1)^n q^{2n}/n!$ on $q^{-1}$ gives
the
notion $\exp(-q^{-2})$.

Finally, since for $f\in\Q$, one has $f=C q C^{-1}$ for some $C\in\U$,
there exists the functional inverse $g=C^{-1} q C\in\Q$ such that
\beq
\label{finv}
\tilde f(g)=q.
\eeq
The $C$ is not
unique, but the non-uniqueness does not affect the notion $g\lt q \rt$, only
that of its conjugate $p_g\lt q,p \rt$.  This means that notions like
$\ln q$ and $q^{1/2}$ are in $\Q$ as inverses in the sense of
(\ref{finv}) to the mappings induced by $e^q$ and $q^2$.  Each of these
could also be defined in terms of formal Laurent expansions about points
other than zero.

One might be concerned about possible branch structure
of the inverses.  The algebra $\U$ is defined over the field of complex
scalars.  Considering elements in $\U$ which involve these explicitly,
one has for example
\beq
(\alpha q)^{1/2}= \alpha^{1/2} q^{1/2}.
\eeq
The branch structure is determined by the functions of the complex scalars.
There is no additional branch structure associated with the generators
themselves.

In the more general case of $f\in\U$, one must define the functional
inverse on the induced mapping of $\M$ to $\M$.  Given $(f,p_f)\in\M$,
there exists $(g,p_g)\in\M$ such that
\beq
(\tilde f(g,p_g), \tilde p_f(g,p_g))=(q,p)
\eeq
If $(f,p_f)=(C q C^{-1}, C p C^{-1})$, then $(g,p_g) =(C^{-1} q C, C^{-1} p
C)$.
This inverse is unique.

This completes the construction of $\U$.  One can see that it is quite
large, consisting essentially of all formal Laurent expansions and
their algebraic and functional inverses.  A description of what is not
included is perhaps more succinct:  one does not allow expressions that involve
distributions, in themselves or in their derivatives.  This is because
the algebraic and functional inverses of a distribution are undefined.
The possibility exists that one might exclude distributions from
$\U$ and $\M$, which constitute the quantum canonical
group and quantum phase space, but allow them as part of an extended class
of notions on which the canonical transformations act.  In this way, one
might treat Hamiltonians that involve distributions.  This will not be
considered here.

\subsection{Representation as Operators}
\label{srep}

The elements of the quantum canonical group $\U$ must be represented by
operators
before they may act on the states of a Hilbert space.  For notational
convenience here and in the sequel, let $(q,p)$ denote all of the phase
space variables $(q_k,p_k)$ ($k=0,\ldots,n$).  The variables $(q,p)$ will
be represented by the operators $(\check{q},\check{p})\equiv (q, -i\d_q)$
acting on functions $\psi(q)$ on configuration space.
These operators are not
to be thought of as self-adjoint operators in the standard inner product
because a Hilbert space has not yet been specified, and in particular
$\psi(q)$ need not be square-integrable.  Functions $C(q,p)\in {\cal U}$
are represented by operators $\check C(\check{q},\check{p})$.  (As
discussed above, to conform with common usage, the convenient fiction is
used from here on that elements $C(q,p)\in\U$ are functions of $(q,p)$
rather than their more precise characterization as the ``notions'' $C\lt
q,p\rt$ expressed in terms of $q$, $p$.)

There is a subtlety in the correspondence of functions in $\cal U$ and
their representations as operators.  The operator $(C^{-1})\check{\vb}$
corresponding to $C^{-1}$ is not always inverse to $\check{C}$ because
the kernels of $\check{C}$ or $(C^{-1})\check{\vb}$ may be non-trivial.
This prevents one from rigorously speaking of the operator
$\check{C}^{-1}$, except when $\check{C}$ is invertible.  Given a function
$\psi$ on which one will apply $\check{C}$, it may be decomposed as a sum
of three parts
$$\psi=\psi_0+\psi_1+\overline{\psi},$$
where $\psi_0\in {\rm ker} \check{C}$; $\psi_1$ in the
pre-image of the kernel of $(C^{-1})\check{\vb}$, i.e., $\check{C}\psi_1\in
{\rm ker}(C^{-1})\check{\vb}$; and the remainder $\overline{\psi}$.
The function
$$\overline{\psi}'=\check{C}\overline{\psi}$$
is then in the class
${\rm Im}\check{C}/{\rm ker}(C^{-1})\check{\vb} $ while $\overline{\psi}\in
{\rm Im}(C^{-1})\check{\vb}/{\rm ker}\check{C}$.
One has the desired properties
\beq
\label{inv1}
(C^{-1})\check{\vb} \check{C} \overline{\psi} =\overline{\psi},
\eeq
and
\beq
\label{inv2}
\check{C} (C^{-1})\check{\vb}\, \overline{\psi}'=\overline{\psi}'.
\eeq
By using $\check{C}$ and $(C^{-1})\check{\vb}$, one has a
definition of the inverse that applies for all
functions that
lie outside the kernels of the respective operators.  As an example, the
action of
$(p^{-1})\check{\vb}$ is indefinite integration modulo an element of ${\rm
ker}\,\check{p}$, that is, modulo an additive constant.

It is believed
that this definition agrees with that of the pseudo-differential
operators, up to infinite smoothing operators\cite{pseudo}.  As will be
seen below, operators corresponding to functions of $p$ are often
evaluated in terms of the Fourier transform of a function of $q$, similar
to the treatment of pseudo-differential operators.

\subsection{Quantum Canonical Transformations}
\label{sqct}

A quantum canonical
transformation is defined\cite{BHJ} as a change of the
phase space variables which preserves the Dirac bracket
\beq
\label{Dbr}
[q,p]=i=[q'(q,p),p'(q,p)].
\eeq
These transformations are generated by an arbitrary complex function
$C(q,p)\in \U$
\beq
CqC^{-1} = q'(q,p) , \quad
CpC^{-1} = p'(q,p).
\eeq
Factor ordering is
built into the definition of the canonical transformation through the ordering
of $C$.  No Hilbert space is mentioned in this definition.

The $C$ producing a given pair $(q',p')\in\M$ is unique (up to a
multiplicative constant).  Suppose that there were two canonical
transformations $C_1$ and $C_2$ implementing the same change of variables
\beqa
C_1 q C_1^{-1} = &q'& = C_2 q C_2^{-1} \\
C_1 p C_1^{-1} = &p'& = C_2 p C_2^{-1}. \non
\eeqa
Multiplying by $C_2^{-1}$ on the left and $C_1$ on the right, one
sees that $C_2^{-1} C_1$ simultaneously commutes with both $q$
and $p$ and therefore must be a multiple of the identity.

The Schr\"odinger operator corresponds to the function
\beq
\sHa(q,p)= p_0 +
H(q_1,\ldots,q_n,p_1,\ldots,p_n,q_0)
\eeq
in $\U$---this will be referred to as the ``Schr\"odinger function.''
The canonical transformation $C$ transforms the Schr\"odinger function
\beq
\label{sHtran}
\sHa'(q,p) = C\sHa(q,p)C^{-1}
= \sHa(Cq C^{-1},Cp C^{-1}).
\eeq
(The action of canonical transformations can be generalized by considering
inhomogeneous transformations $\sHa'=DC\sHa C^{-1}$, $D\in \U$.  This
can be useful in certain applications. $D=1$ is
assumed here; the case with $\check{D}$ invertible requires only minor
modification.)

Solutions of the Schr\"odinger equation $\check{\sHa}'\psi'=0$ are solutions of
$\check{C}\check{\sHa}(C^{-1})\check{\vb}\,\psi'=0$.  If the kernel
of $\check{C}$ is trivial, then
\beq
\label{wftran}
\psi=(C^{-1})\check{\vb}\,\psi'
\eeq
are solutions of the Schr\"odinger equation $\check{\sHa}\psi=0$.
Note that since no inner product has been specified, the transformation
$(C^{-1})\check{\vb}$
acts on all solutions of $\check{\sHa'}$, not merely the normalizable
ones.  When the kernel of $\check{C}$ is non-trivial, the situation is less
simple and requires further discussion.

Before addressing this, consider the uniqueness of the canonical
transformation $C$ between $\sHa$ and $\sHa'$. A symmetry of $\sHa$ is a
transformation $S_\la$ such that $S_\la \sHa S_\la^{-1}=\sHa$. The
symmetries of $\sHa$ form a group. If $\sHa$ has a symmetry $S_{\la}$ and
$\sHa'$ a symmetry $S'_{\mu}$, then the function $S^{\prime\, -1}_{\mu} C
S_{\la}$ is also a canonical transformation from $\sHa$ to $\sHa'$.
Conversely, if $C_a$ and $C_b$ are two canonical transformations from
$\sHa$ to $\sHa'$, then $C_b^{-1}C_a$ is a symmetry of $\sHa$ and $C_a
C_b^{-1}$ is a symmetry of $\sHa'$. This implies that the collection $\cal
C$ of canonical transformations from $\sHa$ to $\sHa'$ are given by one
transformation $C$ between them and the symmetry groups of $\sHa$ and
$\sHa'$.

In constructing the solutions of $\check{\sHa}$ from those of
$\check{\sHa'}$, one must take care when the kernel of $\check{C}$ is
non-trivial.  In this case, there may be solutions $\psi'$ of $\check{\sHa'}$
which by
(\ref{wftran}) produce a $\psi$ which is not a solution of $\check{\sHa}$, but
instead lead to
\beq
\psi''=\check{\sHa}\psi,
\eeq
where
\be
\psi''\in \ker \check{C}.
\ee

To illustrate the problem in a simple case, consider $\sHa=p^3$,
$\sHa'=p^3$.  Clearly, $C=p$ is a canonical transformation, $C\sHa
C^{-1}= \sHa'$.  Consider the solution $\psi'=q^2$ of $\check{\sHa'} \psi'=0$.
By (\ref{wftran}), this gives $\psi=(p^{-1})\check{\vb} \psi'=iq^3/3$.
This is not a solution of $\check{\sHa}$: $\check{\sHa}\psi= -2 \in\ker
\check{C}$.
One has $\check{p} \psi=q^2 =\psi'$ so that $\check{C}$ is invertible on
the solution $\psi$, so this is not the source of the problem.

The problem is simply
that when $\ker\check{C}$ is non-trivial, the transformation
$(C^{-1})\check{\vb}$ can take one outside the
solution space of $\check{\sHa}$.
To deal with this, one must
always check that $\check{\sHa}\psi=0$ for candidate
$\psi=(C^{-1})\check{\vb} \psi'$.  If $\psi$ is not a solution, it has a
decomposition $\psi=\psi_s+ \psi_n$, as the sum of a solution $\psi_s$ and a
non-solution $\psi_n$.  If the
intersection of $\ker\check{C}$ and $\ker(\sHa^{-1})\check{\vb}$ is
empty, then $\check{\sHa}$ is invertible on $\psi_n$.  Thus, one may remove it
from $\psi$ by the projection
$$\psi_s=(1-(\sHa^{-1})\check{\vb}\check{\sHa})\psi.$$
If  $\ker\check{C} \cap \ker(\sHa^{-1})\check{\vb} \not= \emptyset$, one
must work harder.

\subsection{Physical Equivalence and Isometric Transformations}
\label{unieq}

One of the central features of the approach described here is that
canonical transformations need not be unitary.  It is important
however to determine the circumstances under which they are.  Both
in evolution and for establishing the physical equivalence of two
theories, unitary transformations play a key role.  Strictly speaking a
unitary transformation is defined as a linear norm-preserving isomorphism
of a Hilbert space onto itself\cite{DuS}.   For evolution this is fine,
but for physical equivalence it is too restrictive.

Two theories are physically equivalent if
the values of all transition amplitudes are the same in both theories.
This may be true even if the theories are formulated in different Hilbert
spaces, that is, with different inner products.  It would be a physically
and mathematically
sound generalization to extend the definition of a unitary transformation to
this case because this is the spirit in which unitary is meant.
Unfortunately, the secondary definition of a unitary transformation $U$
as one for which $U^\dagger U=1$ follows from the assumption that $U$
maps the Hilbert space onto itself.  This definition is so embedded in the
collective consciousness of physicists that it would be foolish to try to
modify
it.

A linear norm-preserving isomorphism from one Hilbert space onto another
is also known as an isometric transformation\cite{DuS}.  The prudent course is
to
distinguish this from a unitary transformation for which $U^\dagger U=1$.
Then, two theories are physical equivalent if they are
related by an isometric transformation.  The phrase ``unitary equivalence''
is often used as a synonym for ``physical equivalence.''  This usage
while perhaps mildly misleading is acceptable because it tends to call
to mind the
notion that norms are preserved and not a specific formula which is
violated.

To determine the conditions under which a canonical transformation is
isometric,
consider two Schr\"odinger functions $\sHa$ and $\sHa'$ related by a canonical
transformation $C$ (\ref{sHtran}).
For $C$ to define a physical equivalence, it must be an isomorphism
between the Hilbert spaces of $\sHa$ and $\sHa'$ and is therefore
invertible on them.
The solutions
$\psi$ of $\check{\sHa}\psi=0$ are thus given in terms of the
solutions $\psi'$ of
$\check{\sHa}'\psi'=0$ by (\ref{wftran}).

The inner product on the solutions of $\sHa$ has the form
\beqa
\l \phi | \psi \r_{\mu} &\equiv& \l \phi | \check{\mu}(\check{q},\check{p})
| \psi \r_{1} \\
&=& \int d\Sigma\, \phi^*(q) \check{\mu}(\check{q},\check{p}) \psi(q),
\nonumber
\eeqa
where the integration is over spatial configuration space.  Physical
solutions are those which are normalized to either unity or the delta function.
The ``measure density'' $\check{\mu}(\check{q},\check{p})$ may in general
be operator valued and may involve the temporal
variables.  This should not be surprising as the inner product for the
Klein-Gordon equation involves $\check{p}_0$.

When one makes a canonical transformation, in general the measure density
must transform to preserve the norm of states.  This is why one must
consider isometric transformations.
Given the canonical transformation $C$ from $\sHa$
to $\sHa'$, the norm of states is preserved when
\beqa
\l \psi | \psi \r_{\mu} &=&
\l (C^{-1})\check{\vb}\,\psi' |\check{\mu} | (C^{-1})\check{\vb}\,\psi' \r_1
\non \\
&=& \l \psi' | (C^{-1})\check{\vb}\,\vb^\dag1\, \check{\mu} (C^{-1})\check{\vb}
|\psi' \r_1 \nonumber \\
&=& \l \psi' | \psi' \r_{\mu'}. \nonumber
\eeqa
The transformed measure density is
\beq
\label{mtran}
\check{\mu}'(\check{q},\check{p})=(C^{-1})\check{\vb}\,\vb^\dag1\,
\check{\mu}(\check{q},\check{p}) (C^{-1})\check{\vb},
\eeq
or, in $\U$,
\beq
\label{mutran}
\mu'(q,p)=C^{-1\, \dag1} C^{-1} \mu(Cq C^{-1},Cp C^{-1}).
\eeq
Here, $(C^{-1})\check{\vb}\,\vb^\dag1$ is the ``adjoint'' of
$(C^{-1})\check{\vb}$ in the trivial measure density, $\check{\mu}=1$. From
(\ref{mutran}), one sees that the measure transforms as a function on $\U$
multiplied by an inhomogeneous factor.

For isomorphisms of a Hilbert space onto itself, the measure density
does not change. If the measure density is purely a function of the
spatial coordinates, as it usually is in non-relativistic quantum
mechanics, one finds from (\ref{mtran}) the familiar condition for a
unitary transformation: $\check{C}^\dagger \check{C}=1$, where
$\check{C}^\dagger=\check{\mu}^{-1} \check{C}^\dag1 \check{\mu}$ is the adjoint
in the measure density $\check{\mu}$ of the Hilbert space.

The possibility exists for one to redefine the states by absorbing a
factor from the measure-density.  This is an additional canonical
transformation.  For example, if
$\check{\mu}$ and $\check{\mu}'$ are self-adjoint in the trivial measure
density, i.e., $\check{\mu}^\dag1=
\check{\mu}$, $\check{\mu}^{\prime\,\dag1}=\check{\mu}'$, by
redefining the
wavefunction
\beq
\label{resc}
\psi''=\check{\mu}^{-1/2} \check{\mu}^{\prime\,1/2}\psi',
\eeq
the measure density reverts to the original $\check{\mu}$ . The
transformation from $\psi$ to $\psi''$ then preserves the measure density
and so may be unitary. This redefinition can be useful in many
circumstances, but there are times when it is desirable to work with
a transformed measure density.

To summarize, if the Hilbert space of the transformed system $\sHa'$
has the measure given by
(\ref{mtran}) and is isomorphic to the original Hilbert space, the
canonical transformation is isometric.  The quantum theories defined by
$\sHa$ and $\sHa'$ and their Hilbert spaces are then physically
equivalent.  For further discussion, see Ref.~\cite{And1}.

\subsection{Infinitesimal Canonical Transformations}
\label{inftr}

Having explored the general features of quantum canonical
transformations, turn now to consider their connection with classical canonical
transformations and the basis for their explicit construction.
Classically, an infinitesimal canonical transformation is generated by an
(infinitesimal) generating function $F(q,p)$\cite{Gol}.  Associated to this
generating function is the Hamiltonian vector field
\beq
v_F=F\c{p}\d_q- F\c{q} \d_p
\eeq
whose action on a function $u$ on phase space is the infinitesimal
transformation
\beq
\delta_F u=\eps v_F u= -\eps \{F, u\}
\eeq
where $\{F,u\}$ is the classical Poisson bracket. Through the correspondence
between classical and quantum theory in which Poisson brackets times $i$ go
over
into commutator brackets, this canonical transformation can be
expressed in terms of the quantum operator $\exp(i\eps \check{F})$
\beq
e^{i\eps \check{F}}\check{u} e^{-i\eps\check{F}}= \check{u}+i\eps
[\check{F},\check{u}] +O(\eps^2).
\eeq
Because of operator ordering ambiguities in defining the quantum versions
of $\check{F}$ and $\check{u}$, the classical and quantum expressions for the
infinitesimal transformation can differ by higher order terms in $\hbar$
(at the same order of $\eps$).

There are three elementary canonical transformations in one-variable which
have well-known implementations as finite quantum transformations. They are
similarity (gauge) transformations, point canonical (coordinate)
transformations and the interchange of coordinates and momenta. It is
natural to ask what class of transformations is reached by products of
them\cite{LeS}. By finding the algebra generated by the infinitesimal
versions of these transformations, this can be determined. The somewhat
surprising result is that classically they generate the full canonical
algebra. (A partial result in this direction was found by Deenen\cite{Dee}
who did not consider the interchange operation.)

Since the infinitesimal elementary canonical transformations generate the
full classical canonical algebra, in principle, any finite classical
canonical transformation can be decomposed into a product of elementary
canonical transformations. Since each of the elementary canonical
transformations has a quantum implementation as a finite transformation,
the implication is that products of the quantum implementations of the
elementary canonical transformations span the quantum analog of the
classical canonical group. Thus, any quantum canonical transformation can
be decomposed as a product of elementary quantum canonical transformations.
In practice at present, one is limited to finite products of the
transformations. Nevertheless, as will be shown, this is a very powerful
tool for solving problems in quantum mechanics.

It is important to emphasize that the claim is not made that
one can ``quantize'' a classical canonical transformation by
decomposing it into elementary canonical
transformations and then replacing each elementary
transformation with its quantum implementation.  This is not true.  Given
an ordered function of quantum phase space variables and a classical
version obtained by letting the variables commute, if one applies a
sequence of elementary canonical transformations to both, the resulting quantum
and classical expressions will in general not be related by simply
letting the quantum phase space variables commute. This is
consistent with van Hove's theorem\cite{vaH}.

A simple example will cement the point.  The transformation
\beq
p\mapsto p -q^2, \sp q \mapsto q
\eeq
is a canonical transformation both classically and quantum mechanically.
If this transformation is applied to $p^2$, then classically one has
\beq
{\rm classical:}\sp p^2 \mapsto (p-q^2)^2=p^2 -2q^2 p + q^4,
\eeq
while quantum mechanically one has
\beq
{\rm quantum:}\sp p^2 \mapsto (p-q^2)^2=p^2 -2q^2 p +q^4 +2 i q.
\eeq
These are not the same.  The difference is the term $2iq$ which arose
from ordering the latter expression with all of the $p$'s on the right.
One might think that there is some other ordering which would preserve
the correspondence with the classical result.
A consequence of van Hove's theorem\cite{vaH} is that there is no factor
ordering prescription that will preserve the correspondence of the
classical and quantum expressions for all functions on phase space.

The conclusion one draws from this is that one is to put the classical
theory aside
and work exclusively within an ordered quantum theory.  The role of the
classical theory here is solely to motivate the definition of the elementary
canonical transformations. Quantum canonical transformations are
constructed
directly in the quantum theory. The examples below illustrate how they may
be used to solve a theory by transforming to a simpler one whose solution
is known without reference to the classical theory.
In Section~\ref{clvq}, a second example emphasizing
the difference between classical and quantum canonical
transformations is done.

Return to the details of the classical elementary canonical transformations.
Similarity (gauge) transformations are
infinitesimally generated by  $F_S=f(q)$.
Point canonical transformations are infinitesimally generated by
$F_P=f(q)p$.  The discrete transformation $I$
interchanging the role of coordinate and momentum is
\beq
I: p\mapsto -q, \sp
I: q\mapsto p.
\eeq
Using the interchange operator, composite elementary transformations which
are nonlinear in the momentum can be formed.  They are the composite
similarity transformation, infinitesimally generated by
$$ F_{CS}=I F_S I^{-1} =f(p),$$
and the composite point canonical transformation, infinitesimally
generated by
$$F_{CP}=I F_P I^{-1} =-f(p)q.$$
Each of these corresponds to a finite transformation through application
of the interchange operator to the finite forms of the similarity and
point canonical transformations.

The many-variable generalization of the similarity transformation is
straightforward.  It
is infinitesimally generated by $F_{S}=f(q_1,\ldots,q_n)$.
Interchanging  any set of coordinates with their conjugate momenta
gives the composite many-variable similarity transformations.  For example,
in two
variables, one has the infinitesimal generating functions
$F_{CS1}=f(p_1,q_2)$, $F_{CS2}=f(q_1,p_2)$  and $F_{CS12}=f(p_1,p_2)$.

The observation is now made that classically the algebra
generated by
the elementary and composite
elementary transformations is the full canonical algebra
\beq
[v_F,v_G]=-v_{\{F,G\} }.
\eeq
where $F,G$ are arbitrary functions on phase space.  It is to be
expected that
Hamiltonian vector fields will have this algebraic structure.   More
surprising is that the above collection of generating functions
produce a general function on phase space through the Poisson
bracket operation.  This is verified by
taking commutators of the different types of transformations.

Consider the two-variable case---the many-variable case follows similarly.
Introducing the monomial generating functions
\beq
F^{jk}_{nm} = q^{j+1}_{1} q^{k+1}_{2} p^{n+1}_{1} p^{m+1}_{2},
\eeq
where $j,k,n,m\in Z$,
the Poisson bracket of two of these is
\beqa
\{F^{j_1 k_1}_{n_1 m_1},F^{j_2 k_2}_{n_2 m_2}\}&=&
\bigl( (j_2+1)(n_1+1)- (j_1+1)(n_2+1)\bigr)
F^{j_1+j_2\ k_1+k_2+1}_{n_1+n_2\ m_1+
m_2 +1} \\
&&+\bigl( (k_2+1)(m_1+1)-(k_1+1)(m_2+1) \bigr)
F^{j_1+j_2+1\ k_1+k_2}_{n_1+n_2+1\ m_1+m_2}. \nonumber
\eeqa
Inspection shows that a general monomial can be constructed by
beginning with the monomial forms
which generate
the elementary
and composite elementary canonical transformations.

By taking linear
combinations, one can form any function having a Laurent expansion about
some point (not necessarily the origin).  To avoid having to use formal
Laurent expansions about nonzero points for functions
like $q^{1/2}$ or $\ln q$,
greater generality is
obtained by working with generating functions of the form
\beq
F=f_1(q_1) f_2(q_2) g_1(p_1) g_2(p_2).
\eeq
This produces any function which can be represented as a sum of separable
products.


\subsection{Quantum Implementations}
\label{qimpl}

Each of the elementary canonical transformations can be implemented
quantum mechanically as a finite transformation.  Their action is
collected in Fig.~1, and each will be reviewed below.

The interchange of coordinates and momenta
\beq
p\mapsto Ip I^{-1}=-q, \sp
q\mapsto Iq I^{-1}=p.
\eeq
is implemented through
the Fourier transform operator
\beq
\check{I}={1\over (2\pi)^{1/2}} \int_{-\infty}^\infty dq' e^{iq q'}
\eeq
for which it is evident that
\beq
\check{I} q' =-i\d_{q}\check{I}, \hspace{.8cm} \check{I}\check{p'}
=-q\check{I}.
\eeq
The wavefunction is transformed
\beq
\ps1(q)=\check{I}\ps0(q)={1\over (2\pi)^{1/2}}
\int_{-\infty}^\infty dq' e^{iq q'} \ps0(q').
\eeq
The inverse interchange is
\beq
p\mapsto q, \sp q\mapsto -p.
\eeq
It is implemented by the inverse Fourier transform
\beq
(I^{-1})\check{\vb}={1\over (2\pi)^{1/2}} \int_{-\infty}^\infty dq' e^{-iq q'}.
\eeq

The similarity transformation in one-variable is implemented by
\beq
e^{-f(q)},
\eeq
where $f(q)$ is an arbitrary complex function of the coordinate(s). While
the coordinate is unchanged, the momentum transforms
\beq
\label{s}
p \mapsto e^{-f(q)} p e^{f(q)}
 =p -i f\c{q},
\eeq
The wavefunction is transformed
\beq
\ps1(q)=(e^{-f(q)})\check{\vb}\ps0(q).
\eeq
The composite similarity transformation is implemented by applying the
interchange transformation to $e^{-f(q)}$ to exchange coordinates for
momentum in
$f$.  In the one-variable case, the composite similarity transformation
operator is
\beq
e^{-f(p)}=I e^{-f(q)} I^{-1}.
\eeq
It produces the canonical transformation
\beq
\label{cs}
q \mapsto e^{-f(p)} q e^{f(p)} = q +if\c{p} ,
\eeq
while leaving the momentum unchanged.
The wavefunction is transformed
\beq
\ps1(q)=(e^{-f(p)})\check{\vb}\ps0(q)=(I e^{-f(q)} I^{-1})\check{\vb} \ps0(q).
\eeq
Note that the operator corresponding to $e^{-f(p)}$ is defined in terms
of the Fourier transform of a function of $q$, similar to its definition
as a pseudodifferential operator\cite{pseudo}.

In the many-variable case, the function $f$ may involve either the
coordinate or its conjugate momentum for each variable.  Because
variables of different index commute, each variable responds to the (composite)
similarity operator as if
it were a one-variable operator in that variable with the other variables
treated as parameters. Thus,
for each coordinate (momentum) of which $f$ is a
function, the corresponding conjugate momentum (coordinate) is shifted as
in the one-variable case.

For reasons discussed below, it is convenient to represent the finite
point canonical transformation not explicitly as the exponential of the
infinitesimal form, but symbolically as $P_{f(q)}$.
The effect of the point canonical transformation
$P_{f(q)}$ is to implement the change of variables
\beqa
\label{pc}
q &\mapsto& P_{f(q)} q P_{f^{-1}(q)} =f(q), \\
p &\mapsto& P_{f(q)} p P_{f^{-1}(q)} ={1\over f\c{q} } p. \nonumber
\eeqa
The effect of $P_{f(q)}$ on the wavefunction is
\beq
\ps1(q)=\check{P}_{f(q)} \ps0(q)=\ps0(f(q)).
\eeq

The composite point canonical transformation is formed by composition
with the interchange operator
\beq
P_{f(p)}=I P_{f(q)} I^{-1}.
\eeq
It has the effect of making a change of variables on the momentum
\beqa
q &\mapsto&{1\over f\c{p }}q,  \\
p &\mapsto& f(p) . \nonumber
\eeqa
The operator ordering of the transformed $q$ is determined by
the action of the interchange operator on the coordinate point canonical
transformation. The transformed wavefunction is
\beqa
\ps1(q)&=& \check{P}_{f(p)}\ps0(q)=(I P_{f(q)} I^{-1})\check{\vb} \ps0(q) \\
&=& \left( (f\c{p}(p) e^{i f(p) q})\check{\vb} \ps0(q) \right) |_{q=0}. \non
\eeqa

The behavior of the finite transformation obtained by exponentiating the
infinitesimal
generating function of the point canonical transformations,
$ F= g(q)p$, can be computed using the above transformations.
Let $G(q)=\int dq/g(q)$.  For $C=P_{G(q)}$, one has
$Cp C^{-1}= g(q)p$, so that
\beq
e^{iag(q)p}=Ce^{iap} C^{-1}.
\eeq
The action of $e^{iag(q)p}$ on $q$ taking $q\mapsto f(q)$ is then
\beqa
\label{pcf1}
f(q)=\exp(iag(q)p) q \exp(-iag(q)p) &=& Ce^{iap} C^{-1} q C e^{-iap} C^{-1}
\non \\
&=& C e^{iap} G^{-1}(q) e^{-iap} C^{-1}  \\
&=&  C G^{-1}(q +a) C^{-1} \non \\
&=& G^{-1}(G(q)+a) .\non
\eeqa
This result is found by a more laborious method in \cite{Dee}.

As an explicit example, take $g=q^m$.  The
infinitesimal
transformation is then that of the Virasoro generators.  The finite
transformation is (for $m\ne 1$)
\beq
f(q)=e^{iaq^m p}q e^{-iaq^m p}=(q^{1-m} + (1-m)a)^{1/1-m}.
\eeq
while for $m=1$
\beq
f(q)=e^a q.
\eeq
For $m=0,1,2$ the transformations $e^{i\a p}$, $e^{i(\ln\b)qp}$, and
$e^{i\g q^2 p}$ produce translations $q+\a$, scalings $\b q$ and
special conformal transformations $q/(1-\g q)$.  These are well-known to
generate the group $SL(2,C)$.

Eq.~(\ref{pcf1}) gives the finite transformation $q\mapsto f(q)$ produced by
the infinitesimal generating function $ F=g(q)p$.
It can also be read as a functional equation
\beq
G(f(q))=G(q)+a
\eeq
which is to be solved for $G$ given $f$.  Not surprisingly, this equation
does not have a solution for all $f$.  The implication is that not all
point canonical transformations can be expressed as the exponential of an
infinitesimal transformation.  This is an explicit demonstration of the
well-known property of the diffeomorphism group that the exponential map
does not cover a neighborhood of the identity\cite{Mil}.  This is a
property of infinite-dimensional Lie groups and stands in contrast to
the situation in finite-dimensional Lie groups.

A corollary is that the product of the exponentials of two
generators cannot
always be expressed as an exponential of a third generator.  For this
reason, it is generally not useful to express point canonical
transformations in exponential form, but rather to note directly the change
of coordinate they produce.  (As reassurance, it is true that
every point canonical transformation can be expressed as a finite
product of exponentials\cite{Mil}.)

\section{Applications}
\label{Examples}

\subsection{Linear Canonical Transformations}
\label{linct}

The linear canonical transformations form a finite-dimensional
subgroup of all canonical
transformations, and there has been much interest in them in the context of
coherent states\cite{lin,qo}.
As canonical transformations, they can
be constructed
from a product of elementary transformations.

Consider the case of
a single variable. A linear composite similarity transformation
\be
q \mapsto q-i\a p .
\ee
transforms the wavefunction
\be
\ps{a}=(e^{\a p^2/2})\check{\vb} \ps0.
\ee
(A superscript is used to indicate the generation of the transformation.
Subscripts are used to distinguish variables when necessary.)
A linear similarity transformation
\be
p\mapsto p^b -i \b q^b
\ee
makes the change
\be
\ps{b}=(e^{-\b {q}^2/2})\check{\vb}\ps{a}.
\ee
Finally a scaling of the coordinate
\be
p\mapsto {1\over \ga}p, \sp q\mapsto \ga q
\ee
gives
\be
\ps{1}=(e^{i\ln\ga\, q p})\check{\vb}\ps{b}.
\ee
The full transformation is
\beq
\label{sl2}
p\mapsto{1\over \ga}p -i\b \ga q, \sp q\mapsto {-i\a \over \ga}p +
\ga(1-\a\b) q,
\eeq
with
\beq
\label{tr1}
\ps1(q)= (e^{i\ln\ga\, qp}e^{-\b q^2/2} e^{\a p^2/ 2 })\check{\vb} \ps0(q).
\eeq
A general $SL(2,C)\equiv Sp(2,C)$ transformation is of the form
$p=ap' +bq',\ q=cp' +dq'$ where $ad-bc=1$.  This gives the correspondence
$\a=ic/a,\ \b=iab,\ \ga=1/a$ ($a\ne 0$).

By expressing $\ps0$ as the Fourier transform of $\tilde\psi^{(0)}$,
an integral representation
is found for $\ps1$ which does not explicitly involve exponentials
of differential operators
\beq
\label{tr2}
\ps1(q)= {1\over (2\pi)^{1/2}} \int dq'\, e^{i\ga q'q -\b\ga^2 q^2/2
+\a {q'}^2/2} \tilde\psi^{(0)}(q').
\eeq
A related result is given by Moshinsky\cite{Mos}.

The interchange transformation taking $(q,p)\mapsto (p,-q)$
can be constructed similarly as a linear
canonical transformation.  It is found to be
\beq
I=e^{iq^2/2} e^{ip^2/2} e^{i q^2/2}.
\eeq
Using this, one can modify the derivation of (\ref{tr1})
to handle the case where $a=0$.

The operators representing the functions $p^2,\ q^2,\ (qp+pq)/2$ in $\cal U$
generate a realization of the $SL(2,C)$ algebra.  Since $SL(2,C)$ is a
finite-dimensional Lie group, every element of the group in the neighborhood of
the identity can be expressed
as an exponential of an element of the algebra.  As well,
a given linear canonical transformation may be expressed
in many ways as a product of elementary transformations.  Each will give
an expression analogous to (\ref{tr1}) or (\ref{tr2}).

The generalization to many-variables is straightforward.  The
group of linear canonical transformations is $Sp(2n,C)$, and
a realization of it is found from the linear similarity,
composite similarity and scaling transformations.  Realizations of
other finite-dimensional Lie groups in terms of canonical
transformations are found by
treating them as subgroups of $Sp(2n,C)$.

By expressing the coordinates and momenta in terms of harmonic oscillator
creation and
annihilation operators, one finds
the expressions for the action of linear canonical transformations on
coherent states\cite{lin}.  These are useful for handling squeezed states
in quantum optics\cite{qo}.

\subsection{Non-relativistic Free Particle}
\label{sfree}

The non-relativistic free particle is an easily solved problem which
doesn't require any sophisticated machinery.  It may however serve to
illustrate a number of features of the use of canonical transformations,
and, for this reason, it will be treated somewhat exhaustively.

The free particle Schr\"odinger function is
\beq
\sH0= p_0+ p^{2}.
\eeq
It may be immediately trivialized
\beq
\sH{1}=p_0
\eeq
by the two-variable similarity transformation
\beq
p_0 \mapsto p_0 -p^2, \sp q\mapsto  q +2p q_0.
\eeq
(Variables left unchanged by a transformation will be suppressed.)
The original wavefunction is given in terms of the transformed one by
\beq
\label{wf0}
\ps0(q, q_0)=(e^{-ip^2  q_0})\check{\vb} \ps{1}(q).
\eeq
where the solution of the trivialized Schr\"odinger equation
$\check{\sH{1}}\ps{1}=0$ is any $q_0$-independent function.
This formula
is just the formal expression for the evolution of an
initial wavefunction in terms of the exponential of the Hamiltonian.
In general,
this formal result is insufficiently explicit, but for the free particle,
it can be used to find more useful forms of the wavefunction.

For example, if the initial wavefunction is taken to be a plane wave
$\ps{1}=\exp(ikq)$, one finds the plane wave stationary solution
\beq
\ps0=e^{ikq-ik^2 q_0}.
\eeq
If the initial wavefunction is a delta function at $x$, then using the
Fourier integral representation of the delta function, the wavefunction
is
\beq
\label{Fdel}
\ps0=(e^{-ip^2  q_0/2})\check{\vb}{1\over 2\pi}
\int_{-\infty}^\infty dq' e^{i(q-x)q'}.
\eeq
Acting with the operator inside the integral and integrating the
resulting Gaussian gives
\beq
\label{Gf}
\ps0=(4\pi i q_0)^{-1/2} e^{i(q-x)^2/4 q_0}.
\eeq
This is the free particle Green's function.

It is clear that the nature
of the wavefunction depends on the initial wavefunction used to generate
it.  This obvious comment is important to bear in mind because the
``natural'' solution of a transformed problem will not always correspond
to the desired solution.  To see this, consider a second approach to the
free particle.
The interchange transformation
\be
p \mapsto - q,\sp
q \mapsto p
\ee
is equivalent to taking the Fourier transform of the original
Schr\"odinger equation and gives
\beq
\sH{a}= p_0+ q^2.
\eeq
The original wavefunction is given by the inverse Fourier transform
\beq
\ps{0}{}(q)={1\over (2\pi)^{1/2}} \int_{-\infty}^\infty dq'
e^{-iq q'} \ps{a}(q').
\eeq

The new equation $\check{\sH{a}}\ps{a}=0$ can be solved ``naturally''
in two ways.
The first is simply to integrate with respect
to $ q_0$.  This gives
\beq
\label{fpwa}
\ps{a}=f(q)e^{-iq^2  q_0}
\eeq
where $f(q)$ is an arbitrary $q_0$-independent function that
arises as an integration constant.  One finds the wavefunction
\beq
\label{wf1}
\ps0= {1\over (2\pi)^{1/2}} \int_{-\infty}^\infty dq'
e^{-iq q'-i{q'}^2  q_0} f(q').
\eeq
This is a less familiar form of the wavefunction---it is of course just
a momentum space representation---and it is perhaps not immediately
obvious how to obtain the solutions above.  Inspection shows that if
$f=\delta(q'+k)$, one finds the plane wave stationary solution.
Alternatively, if $f$ is taken to be
\beq
f={1\over (2\pi)^{1/2}} e^{ix q'},
\eeq
then, at $q_0=0$, one has $\ps0=\delta(q-x)$ and, evaluating the
integral, one finds again the Green's function (\ref{Gf}).

The second ``natural'' approach is to separate variables
\beq
\ps{a}=\ph{a}(q)e^{-ik^2 q_0}.
\eeq
This results in the equation
\beq
(q^2 -k^2)\ph{a}=0,
\eeq
which has as its solution
\beq
\ph{a}{}=\delta(q -k)
\eeq
($k$ can have either sign).
Now inverting the interchange operation gives the familiar plane wave
solution
\beq
\ps0={1\over (2\pi)^{1/2}} e^{ ik q-ik^2  q_0}.
\eeq
The Green's function solution is however no longer obtainable.

The first of these approaches can itself be implemented as a canonical
transformation.  The similarity transformation
\beq
p_0 \mapsto p_0 -q^2 ,\sp  p \mapsto p-2q q_0
\eeq
trivializes the Schr\"odinger function
\be
\sH{1}=p_0
\ee
and gives the wavefunction
\beq
\ps{a}=(e^{-q^2 q_0})\check{\vb}\ps{1}.
\eeq
The wavefunction $\ps{1}$ is any $ q_0$-independent function.  The
result for the original wavefunction is then (\ref{wf1}).

The second approach of separation of variables is not so much a canonical
transformation as a realization of the assertion that the solution
space of the
Schr\"odinger operator has a product structure.

\subsection{Harmonic Oscillator}

The harmonic oscillator is the paradigmatic problem in quantum mechanics,
and it is a test piece for any method. Its solution by canonical
transformation reemphasizes how the form of a solution is affected by the
details of evaluating the product of operators representing the canonical
transformation.  Also, it is observed that more than one canonical
transformation to triviality is needed to obtain both independent
solutions of the original Hamiltonian.

The Hamiltonian for the harmonic oscillator is
\beq
\label{ho}
\H0=p^2 +\w^2 q^2.
\eeq
A similarity transformation $\exp(\w q^2/2)$, taking
\be
p \mapsto p +i\w q ,
\ee
will cancel the quadratic term in the coordinate leaving
\be
\H{a} = p^2 +2i\w q p +\w.
\ee
This is recognized as corresponding to the operator for the Hermite
polynomials.
The transformation $\exp(\w q^2/2)$ is real and therefore not unitary.
It is however an isometric transformation.  From (\ref{mtran}), the
measure density in the transformed inner product is $\check{\mu}^{(a)}=
e^{-\w q^2}$, which is the measure density in which the operator for the
Hermite polynomials is self-adjoint.

Since solution by power series expansion is not a canonical transformation,
so much as a
method of approximation, additional transformations are needed.
The composite similarity transformation $\exp(-p^2/4\w)$ takes
\be
q \mapsto q +ip/2\w
\ee
and cancels the quadratic term in the momentum, giving the Hamiltonian
\beq
\H{b}= 2i\w q p +\w.
\eeq
Finally, the point canonical transformation $P_{e^q}$, taking
\be
p \mapsto e^{-q}p, \sp
q\mapsto e^{q},
\ee
transforms the Hamiltonian to action-angle form
\beq
\H1= 2i\w p +\w.
\eeq

In terms of the full Schr\"odinger function, this is
\beq
\sH1=p_0+ 2i\w p +\w.
\eeq
A final two-variable similarity transformation produced by
$\exp(-\w(2 p-i) q_0)$,
\beqa
p_0\mapsto p_0- 2i\w p-\w , \sp
q \mapsto q+ 2 i\w q_0, \nonumber
\eeqa
trivializes this, leaving
\beq
\sH2=p_0.
\eeq

The wavefunction $\ps1$ is given in terms of $\ps0$ by
\be
\ps1(q)= (P_{e^q} e^{-p^2/4\w} e^{\w q^2/2})\check{\vb} \ps0(q),
\ee
which may be inverted to find
\beq
\ps0(q)=( e^{-\w q^2/2} e^{p^2/4\w} P_{\ln q})\check{\vb} \ps1(q).
\eeq
{}From the eigenfunctions of $\check{H}^{(1)}$
\be
\ps1_n(q)=e^{nq-i(2n+1)\w  q_0},
\ee
one has
\beq
\ps0_n(q)=e^{-\w q^2/2}e^{-(\d_q)^2/4\w} q^n e^{-i(2n+1)\w
 q_0}.
\eeq
This is the correct (unnormalized) harmonic oscillator eigenfunction. This
formula is valid for complex $n$. Requiring that the wavefunction be
normalizable fixes $n$ to be a non-negative real integer.  For other $n$,
one finds an infinite power series in $1/q$ which is divergent at
$q=0$.

As remarked above, $\check{H}^{(a)}$ is the Hamiltonian whose solutions are the
Hermite polynomials.  This implies
\beq
\label{herep}
H_n(q) \propto e^{-(\d_q)^2/4} q^n
\eeq
Given this form, it is immediate that $\d_q$ is the lowering operator
\beq
\d_q H_n(q) \propto n H_{n-1}(q).
\eeq
This result could also be found by observing that $p$ is a canonical
transformation from $\H{a}$ to $\H{a}+2\w$.  This transformation has a
non-trivial kernel:  the constant solution $H_0(q)$ is annihilated by the
transformation.  This means that $p$ is not an isomorphism from the
Hilbert space of $\H{a}$ to that of $\H{a}+2\w$.  It is therefore not a
unitary transformation.  Despite being non-unitary, the importance of the
lowering operator cannot be denied.

Furthermore, if one computes the transformed measure density produced by
$p$, one finds it is operator valued and not the familiar coordinate
dependent expression. By absorbing the square-root of this operator valued
measure density into the wavefunction by (\ref{resc}), one can recover the
standard inner product. This produces an $n$-dependent renormalization of
the wavefunctions. This renormalization factor is the $n$-dependent factor
that is present in the lowering operator relating normalized harmonic
oscillator wavefunctions.

The representation (\ref{herep}) of the Hermite polynomials is unfamiliar
because of the operator produced by
the composite similarity transformation between $\H{a}$ and
$\H{b}$.
If this transformation is decomposed into
elementary
canonical transformations, direct evaluation leads to
the more familiar Rodriques' formula for the Hermite polynomials.
The decomposed transformation is
\beq
\ps{a}(q)={1\over 2\pi} \int_{-\infty}^\infty d\bq e^{i\bq q}
e^{\bq^2/4\w} \int_{-\infty}^\infty d q' e^{-i\bq q'}
{q'}^n.
\eeq
This may be evaluated by first rewriting
${q'}^n$
as $(i\d_\bq)^n$ acting on the exponential $\exp(-i\bq q')$.
It may be extracted from the $q'$ integral which then
gives $\delta(\bq)$.
Integrating by parts $n$ times transfers the $i\d_\bq$ operators to act on
the remaining exponential terms
\beq
\ps{a}(q)= \int_{-\infty}^\infty d\bq (-i\d_\bq)^n
e^{i\bq q+\bq^2/4\w}
\delta(\bq).
\eeq

Completing the square of the argument of the exponential gives
$${1\over 4\w}(\bq+2i\w q)^2 +\w q^2,$$
from which the purely $q$ part can be extracted from the integral.
The $-i\d_\bq$ derivatives act equivalently to $-\d_q/2\w$ derivatives,
and after converting them, they can be removed from the integral.
This leaves a Gaussian integrated against a delta function which is
immediately evaluated.  The result is
\beq
\ps{a}(q)=e^{\w q^2}({-\d_q\over 2\w})^n e^{-\w q^2}.
\eeq
This is proportional to the Rodrigues' formula for the Hermite polynomials
\beq
H_n(\xi)=e^{\xi^2}(-\d_\xi)^n e^{-\xi^2}.
\eeq
{}From this form, it is immediate that
$e^{\xi^2}(-\d_\xi)e^{-\xi^2} =-\d_\xi +2\xi$ is
the raising operator.

There is a second linearly-independent solution of the harmonic
oscillator which was not obtained by this canonical transformation.  This
solution is not normalizable, but from the standpoint of simply solving
the differential equation, this is not important.  A second canonical
transformation which trivializes the Schr\"odinger function in a different way
produces the other solution.  This solution is easily found by observing
that the harmonic oscillator Hamiltonian is symmetric under $\w\mapsto
-\w$.  Making this symmetry transformation on the canonical
transformation above gives the transformation to the action-angle
Hamiltonian
\beq
\H{1'}= -2i\w p -\w.
\eeq
The Schr\"odinger function could be trivialized by a two-variable similarity
transformation, but this is unnecessary.

The wavefunction $\ps{0'}$ is given in terms of $\ps{1'}$ by
\beq
\ps{0'}(q)= (e^{\w q^2/2} e^{-p^2/4\w} P_{\ln q})\check{\vb} \ps{1'}(q).
\eeq
{}From the eigenfunctions of $\H{1'}$
\be
\ps{1'}_n=e^{-(n+1)q-i(2n+1)\w  q_0},
\ee
one has
\beq
\ps{0'}_n(q)=e^{\w q^2/2}e^{(\d_q)^2/4\w} q^{-(n+1)}
e^{-i(2n+1)\w  q_0}.
\eeq
These are clearly not normalizable for any $n$.

The problem of the inverted harmonic oscillator with Hamiltonian
\beq
\H0= p^2 -\w^2 q^2
\eeq
can be solved by the above canonical transformations after $\w$ is
replaced by $i\w$.  This is a scattering problem, so both independent
solutions are delta-function normalizable, with no quantization of
$n$.  This emphasizes the importance
of both canonical transformations to triviality.

\subsection{Harmonic Oscillator Propagator}
\label{shoprop}


The propagator for the harmonic oscillator is formally given by
\beq
\label{prop1}
K(q,q_0|a,0)=(e^{-i\H0 q_0})\check{\vb} \delta(q-a).
\eeq
This can be evaluated by canonical transformations.  The Hamiltonian for
the harmonic oscillator
is transformed to the Hamiltonian in action-angle form
\beq
\H1=i2\w p+\w,
\eeq
by the canonical transformation
\beq
\H1=C \H0 C^{-1},
\eeq
where
\beq
C=P_{e^q} e^{-p^2/4\w} e^{\w q^2/2}.
\eeq
This means that the propagator can be
expressed by a product of elementary canonical transformations
\begin{eqnarray}
\label{prop2}
K(q,q_0|a,0) &=& (C^{-1}e^{-i\H1 q_0}C)\check{\vb} \delta(q-a) \\
&=& (e^{-\w q^2/2} e^{p^2/4\w} P_{\ln q} e^{(2\w p-i\w)q_0} P_{e^q}
e^{-p^2/4\w} e^{\w q^2/2})\check{\vb} \delta(q-a). \nonumber
\end{eqnarray}

Evaluate (\ref{prop2}) starting from the right.  The first product is
$$F_1= e^{\w q^2/2} \delta(q-a)= e^{\w a^2/2} \delta(q-a). $$
The next product is
$$F_2= (e^{-p^2/4\w})\check{\vb}F_1.$$
This can be evaluated in two ways.  One can recognize that
$(e^{-p^2/4\w})\check{\vb}$ is the unitary operator generating free-particle
($H=p^2$)
propagation for time $t=-i/4\w$ and use the
propagator
\beq
K_{free}(q,t|q',0)=(4\pi i t)^{-1/2} e^{i(q-q')^2/4 t}.
\eeq
Or, one can express $e^{-p^2/4\w}$ as
$$e^{-p^2/4\w} =I e^{-q^2/4\w} I^{-1},$$
and, by evaluating the action of this on
$\delta(q-a)$, derive the free particle propagator.  This is essentially
what is done in (\ref{Fdel}).  The result is that
$$F_2=(\w/\pi)^{1/2}e^{-\w(q-a)^2 +\w a^2/2}.$$

The third product is the application of the point canonical transformation
\begin{eqnarray*}
F_3 &=& \check{P}_{e^q}F_2 \\
& =&(\w/\pi)^{1/2}\exp(-\w(e^q-a)^2 +\w a^2/2).
\end{eqnarray*}
The fourth product is a translation of $q$ by $-i2\w q_0$
\begin{eqnarray*}
F_4 &=& (e^{(2\w p-i\w)q_0})\check{\vb} F_3 \\
& =&(\w/\pi)^{1/2}\exp(-\w(e^{q-i2\w q_0}-a)^2 +\w a^2/2-i\w q_0).
\end{eqnarray*}
This is followed by another point canonical transformation
\begin{eqnarray*}
F_5 &=& \check{P}_{\ln q}F_4 \\
& =&(\w/\pi)^{1/2}\exp(-\w(q e^{-i2\w q_0}-a)^2 +\w a^2/2-i\w q_0).
\end{eqnarray*}
A second free particle evolution gives
\begin{eqnarray*}
F_6 &=& (e^{p^2/4\w})\check{\vb} F_5 \\
&=& i\w/\pi \int dq' e^{\w (q-q')^2}\exp(-\w(q' e^{-i2\w q_0}-a)^2 +\w a^2/2
-i\w q_0) \\
&=& \left({ \w \over 2\pi i \sin 2\w q_0}\right)^{1/2}
\exp(i{\w a^2 \over 2} \cot 2\w q_0 -i{\w a q\over \sin 2\w q_0}
-{\w q^2 \over e^{4i\w q_0} -1 })
\end{eqnarray*}

Lastly, multiplying by $e^{-\w q^2/2}$ gives the final result
\beq
K(q,q_0|a,0) = \left({ \w \over 2\pi i\sin 2\w q_0}\right)^{1/2}
\exp(i{\w  \over 2\sin 2\w q_0}\biggl( (a^2 +q^2) \cos 2\w q_0 -2 a
q\biggr) ).
\eeq
This is the familiar Green's function for the harmonic oscillator.

\subsection{Time-Dependent Harmonic Oscillator}
\label{stdho}

The time-dependent harmonic oscillator can also be solved by canonical
transformation. This has been done previously with a different sequence
of canonical transformations by Brown\cite{Bro}.  Here, the approach will
be to parallel the solution of the time-independent harmonic oscillator
in trivializing
the Schr\"odinger function.  This emphasizes the connection between the
time-dependent and time-independent problems.  It is conjectured that
the parallel structure shown here carries over to time-dependent
versions of other exactly
soluble problems.

One begins with the Schr\"odinger function
\beq
\sH0 =p_0 +p^2 +\w^2(q_0) q^2
\eeq
where the angular frequency $\w(q_0)$ is a function of time.
The quadratic term in the coordinate is cancelled by making a
two-variable similarity transformation,
$\exp(-f(q_0) q^2/2)$.  This gives the Schr\"odinger function
\beq
\sH{a}=p_0 +p^2 -2 i f q p
+(2 \w^2-i f \c{q_0} -2 f^2) q^2/2
-f.
\eeq
The condition that the quadratic potential be cancelled is
\beq
i f\c{q_0} +2 f^2 =2 \w^2.
\eeq
This Ricatti equation is linearized by the substitution
\beq
f=i{\phi \c{q_0}\over 2\phi}
\eeq
which gives
\beq
\phi \c{ q_0 q_0} = -4 \w^2 \phi.
\eeq
To go further requires the specific time-dependence of $\w(q_0)$.

The next step is to cancel the quadratic term in the momentum with
a second two-variable similarity transformation,
$\exp(-g(q_0) p^2/2)$.  This gives the Schr\"odinger function
\beq
\sH{b}=p_0 +(2-i g\c{q_0} + 4 f g)p^2/2
-2 i f q p -f
\eeq
with the condition
\beq
i g\c{q_0} - 4 f g=2.
\eeq
This equation can be integrated to find
\beq
g= \exp(-i 4 \int f dq_0) \int -2i \exp(i 4\int f dq_0) d q_0.
\eeq

The Schr\"odinger function  is then
\be
\sH{b}=p_0  -2 i f q p -f.
\ee
The coordinate change
\beq
p \mapsto e^{-q}p, \sp
q \mapsto e^{q}
\eeq
eliminates the coordinate from the Schr\"odinger function
\beq
\sH{c} = p_0 - 2i f p -f .
\eeq
Finally, the Schr\"odinger function is trivialized to
\beq
\sH1= p_0
\eeq
by the composite two-variable shift
$$\exp(-(-2 p +i) \int f  dq_0). $$

Since $\ps1$ is any $q_0$-independent function, this gives the
result
\beq
\ps0(q,q_0)=(e^{f(q_0) q^2/2} e^{g(q_0) p^2/2} P_{\ln q}
e^{(-2 p +i) \int f( q_0)  d q_0})\check{\vb} \ps1(q).
\eeq
The time-independent result is recovered when $f=-\w$ and
$\ps1(q)=e^{nq}$.

\subsection{Intertwining}
\label{sio}

Intertwining is the existence of an operator $D$ which transforms between
two operators
\beq
\label{ior2}
\H{1}D=D\H{0}.
\eeq
It is a powerful method for solving differential equations because it
gives the solutions of $\H{1}$ in terms of those of $\H{0}$.
All of the recursion relations and Rodrigues' formulae for the
classical special functions can be understood as a consequence of
intertwining.

The most widely known example of intertwining
is that of the Darboux transformation between two Hamiltonian operators
in potential-form\cite{And4}
\beq
\H{i}=p^2 +V_i(q).
\eeq
One would like to know what potentials can be reached from a given one by
an intertwining transformation (\ref{ior2}). In the usual approach,
an ansatz is made
that the intertwining operator $D$ is a first order differential operator,
and the operator is
constructed by requiring that the intertwining relation (\ref{ior2})  be
satisfied. It
is satisfactory to begin with first order operators because the
intertwining transformation can be iterated.

Since the intertwining operator is a canonical transformation, it can be
constructed as a sequence of elementary canonical
transformations.  The construction is interesting because it involves a
transformation in which operator ordering makes the the quantum case
simpler than the classical. This is the source of the power of the differential
operator ansatz, and it is the first example of a transformation which
favors the quantum problem over the classical.

Begin with a Hamiltonian with potential
\beq
\label{H0}
\H0=p^2 +V_0(q).
\eeq
The potential may be cancelled by making a similarity transformation
\be
p \mapsto p -ig(q)
\ee
This gives the Hamiltonian
\beq
\H{a} = p^2 -2i g p -\la
\eeq
together with the Ricatti equation
\beq
\label{Ric0}
g\c{q}+g^2 =V_0+\la.
\eeq
The transformed wavefunction is
\beq
\ps{a}=(e^{-\int g dq})\check{\vb} \ps0.
\eeq

A composite similarity transformation is made on the coordinate with $p$
\beq
\label{dio}
q \mapsto pq {1\over p}= q -{i\over p}.
\eeq
This is the key step in an intertwining transformation.
It has the very interesting property
\beq
g(q - {i\over p}) = g(p q {1\over p})=g(q)-ig(q) \c{q}{1\over p }.
\eeq
Note that only the first term in the Taylor expansion of $g$
appears---classically the full Taylor expansion would have arisen.
After the transformation, the Hamiltonian becomes
\beq
\H{b}=p^2 -2i g p-2 g\c{q}-\la.
\eeq
and the transformed wavefunction is
\beq
\ps{b}(q)=(e^{\int dp/p})\check{\vb} \ps{a}(q)=
\check{p} \ps{a}(q).
\eeq
Because the kernel of $\check{p}$ is non-trivial, this
transformation may be non-unitary for the reasons discussed above in the
context of the lowering operator for the Hermite polynomials.
Nevertheless, it is useful.

The transformation
\be
p \mapsto p +i g(q)
\ee
cancels the term linear in the momentum giving the Hamiltonian
\beq
\H1=p^2 +V_1(q),
\eeq
with the new potential
\beq
\label{Ric1}
V_1=-g\c{q}+ g^2-\la.
\eeq
The transformed wavefunction is
\beq
\ps1(q)=(e^{\int g dq})\check{\vb} \ps{b}(q).
\eeq
In terms of the original wavefunction, this is
\beq
\label{io1}
\ps1(q)=(e^{\int g dq}p e^{-\int g dq})\check{\vb} \ps0(q)=
 -i(\d_q -g)\ps0(q).
\eeq

Comparing (\ref{Ric0}) and (\ref{Ric1}), the change in potential is
\beq
V_1-V_0 =-2 g\c{q}.
\eeq
The Ricatti equation (\ref{Ric0}) [or (\ref{Ric1})] can
be solved to find that $g$ is given by the logarithmic
derivative of an eigenfunction of $\H0$ (or the negative logarithmic
derivative of an eigenfunction of $\H1$) with eigenvalue $\la$.
This is the standard result from intertwining\cite{And4}.

If one inverts (\ref{io1}) to obtain $\ps0$ in terms of $\ps1$,
one obtains the integral operator expression
\beq
\label{io1a}
\ps0(q)=(e^{\int g dq}p^{-1} e^{-\int g dq})\check{\vb} \ps1(q).
\eeq
To obtain a differential operator
relation, one may note that taking $g\rightarrow -g$ interchanges
$V_0$ and $V_1$.  This implies from (\ref{io1}) that
\beq
\label{io2}
\ps0(q) = -i( \d_{q} +g) \ps1(q).
\eeq

Alternatively, a different sequence of canonical transformations can
be used which give an integral operator relating $\ps0$ to $\ps1$ which
becomes a differential operator upon inversion.  Beginning from $\H0$
(\ref{H0}), the one-variable shift $p \mapsto p +i g(q)$
gives the Hamiltonian
\beq
\H{a}=p^2 + 2i p g -\la
\eeq
where $V_0$ satisfies (\ref{Ric0}).
Note that $p$ has been ordered on the left of $g$.  This is to
facilitate the transformation
\be
q\mapsto {1\over p} q p= q + {i\over p}.
\ee
This leads to the transformed Hamiltonian
\beq
\H{b} = p^2 +2i p g - 2 g\c{q} -\la.
\eeq
A similarity transformation $p\mapsto p -i g(q)$ cancels the linear momentum
term, leaving the final Hamiltonian
\beq
\H1= p^2 +V_1,
\eeq
where $V_1$ is given by (\ref{Ric1}).
The final wavefunction in terms of the original is
\beq
\ps1(q) =(e^{-\int g dq}p^{-1}e^{\int g dq})\check{\vb} \ps0(q).
\eeq
Inverting this gives the expected differential relation (\ref{io2}).

The operator in (\ref{io2}) annihilates the wavefunction $e^{-\int g
dq}$.  If this wavefunction is in the Hilbert space of $\H1$, the
transformation is not an isomorphism, and the operator is non-unitary.
There is a solution of $\H0$ corresponding to this wavefunction. To obtain
it, one applies the canonical transformation (\ref{io1a}) to
$e^{-\int g dq}$.  This gives a familiar integral expression\cite{Ince} for the
solution complementary to $e^{\int q dq}$, which can be checked to be the
other solution of $\check{H}^{(0)}$ with zero eigenvalue.

The method of intertwining has been realized in terms of canonical
transformations.  This means that all of the problems which can be solved
by intertwining can be solved with canonical transformations.
This includes all problems which are essentially
hypergeometric, confluent hypergeometric, or one of their generalizations.
Furthermore, the Rodrigues' and
differential recursion formulae for the special functions may all be obtained
from canonical transformations of this kind.

\subsection{Classical vs. Quantum Transformations}
\label{clvq}

The distinct behaviors classically and quantum mechanically
of a sequence of elementary canonical transformations
were discussed in Section~\ref{inftr}.
A different perspective on this will given by considering the classical
and quantum transformations between two Hamiltonians whose classical forms
are the same as their quantum forms in a natural ordering.   The point
will be to show that a different sequence of elementary transformations
is needed to implement the transformation classically and quantum
mechanically.  Furthermore, it will be shown that the accumulated
transformation is a highly non-trivial factor ordering of the classical
transformation.  This indicates that attempts to solve a quantum problem
by factor
ordering the classical canonical transformation will generally be in vain.
Only
in very special cases will the quantum ordering be sufficiently simple,
e.g., polynomial, that one could hope to find it by hand.  Constructing
the sequence of elementary transformations which solve the problem
quantum mechanically is a more fruitful course.

The example is the transformation between the Hamiltonians
\beq
\H0=p^2 +e^{2q} \mapsto \H1=p^2.
\eeq
These Hamiltonians have the same form classically and quantum
mechanically if they are ordered as written.
Quantum mechanically, the solutions of $\check{H}^{(0)}$ are Bessel
functions.  The
transformation to $\H1$ would give their construction in terms of plane
waves. This will not be the emphasis here.  Rather, the focus will be on the
difference between the sequence of elementary transformations which
implement the transformation classically and quantum mechanically.

Consider the problem classically.  The transformation
\beq
q\mapsto \ln q,\sp p\mapsto q p,
\eeq
simplifies the potential
\beq
\H{a}_{cl}=q^2 p^2 + q^2.
\eeq
An interchange
\beq
(q,p)\mapsto (p,-q)
\eeq
gives
\beq
\H{b}_{cl}=(1+q^2) p^2.
\eeq
Finally, the change of variables
\beq
\label{qsinh}
q\mapsto \sinh q, \sp p\mapsto {1\over \cosh q} p,
\eeq
gives
\beq
\H1_{cl}=p^2.
\eeq
Denoting the phase space variables in the final equation by $(q',p')$,
the full accumulated transformation in terms of the initial variables is
\beq
\label{cltr}
q'=\arcsinh(-e^{-q}p), \sp p'=(p^2+e^{2q})^{1/2}.
\eeq

Now consider the problem quantum mechanically.  The first two transformations
are the same and give
\beq
\H{a}_{qu}=q^2 p^2 -i q p + q^2,
\eeq
and
\beq
\H{b}_{qu}=p^2 (1+q^2) + i p q.
\eeq
The transformation
\beq
q\mapsto {1\over p}q p= q+ {i\over p}
\eeq
leads to
\beq
\H{c}_{qu}=(1+q^2) p^2 -i q p.
\eeq
Again, the point canonical transformation (\ref{qsinh}) ends the process
\beq
\H1_{qu}=p^2.
\eeq

Note that the sequences of elementary transformations which perform the
full transformation classically and quantum mechanically are similar, but
different.  An extra transformation is needed quantum mechanically to
cancel a term that arose from ordering.  The full quantum transformation is
\beqa
q' &=&\arcsinh( -e^{-q}p) - {i\over \bigl(1+(e^{-q}p)^2 \bigr)^{1/2}}e^{-q},
\\
p' &=& e^q \bigl(1+(e^{-q}p)^2 \bigr)^{1/2}. \non
\eeqa
The functions appearing in these expressions are defined in terms of
their power series expansions.  Clearly, the ordering is non-trivial, and
while the correspondence with (\ref{cltr}) is evident, it would have been
difficult to discover by hand.

This emphasizes the point that classical and quantum canonical
transformations are not simply related.  It is more fruitful to construct
the quantum canonical transformation directly than to attempt to factor
order a classical transformation.

\subsection{Particle on $2n+1$-sphere}
\label{nsph}

As a final example, consider the radial Hamiltonian for a particle
propagating on an $2n+1$-dimensional sphere
\beq
\H0=p^2 -2ni \cot q\,p.
\eeq
Its solution by canonical transformations illustrates the use of the
intertwining canonical transformation. First make the point
canonical transformation
\be
{-1\over \sin q} p \mapsto p, \sp \cos q \mapsto q.
\ee
This is stated in inverse form, contrary to the convention followed
above.  This is often useful with point canonical transformations because
it is more intuitive when looking for changes of variable to simplify an
equation.  The transformed Hamiltonian is found to be
\beq
\H{a}= (1-q^2) p^2 + (2n+1) i q p.
\eeq
This is recognized as corresponding with the equation for the
Gegenbauer polynomials.

The intertwining transformation $p^{-n}$ taking
\be
q \mapsto q +{n i\over p}
\ee
can now be used to cancel the $n$-dependence of the Hamiltonian.
Noting that
\beq
q^2\mapsto ({1\over p^n} q p^n)^2=
q^2 +2n i q
{1\over p} - {n^2+n \over p^2},
\eeq
one finds
\beq
\H{b}=(1-q^2) p^2 + i q p -n^2.
\eeq
Clearly, it would have been possible to shift $n$ by any amount:  this
gives the relation between Gegenbauer polynomials of different
$n$.

Finally, undoing the original point canonical transformation
\be
p \mapsto {-1\over \sin q} p, \sp
q \mapsto \cos q,
\ee
gives the free-particle Hamiltonian
\beq
\H1=p^2 -n^2.
\eeq
Because the physical problem was that of a free-particle on a sphere,
this is the free-particle on a circle.  The spectrum of $\H1$ is
discrete, and
the constant shift produces a time-dependent phase factor $e^{in^2 q_0}$
relative to the usual free-particle eigenfunctions on the circle
\beq
\ps1_m=\cos mq\, e^{-im^2  q_0}.
\eeq
(The other independent solution is found by using $\sin mq$.)
The original wavefunctions are given in terms of the free-particle
eigenfunctions by
\beq
\ps0_m=(P_{\cos q}p^n P_{\arccos q})\check{\vb} \ps1_{m+n}(q)=
\left( {i\over \sin q}\d_q
\right)^n \cos (m+n)q\, e^{-i(m^2+2m n)  q_0}.
\eeq
The indexing is determined by the condition that $m=0$
corresponds to
the normalizable solution with lowest energy.
This agrees with the result obtained by the
intertwining method\cite{And3} and is recognized as a formula for the
Gegenbauer polynomials $c^{(n)}_m(\cos q)$.

In principle, this result is valid for
real $n$.  For $n$ non-integer, one requires an
integral representation of the fractional differential operator.  It is
likely that this can be constructed by manipulating the definition of the
composite similarity transformation in terms of the Fourier transform of
an ordinary similarity transformation.  This would be analogous to the
discussion of the origin of the Rodrigues' formula for the Hermite
polynomials.  There are subtleties involving endpoints of the
integrals which are beyond the scope of this paper.  I hope to return to
this in a later work.

\section{Conclusion}

It has been shown how, using a few elementary canonical transformations
which have quantum mechanical implementations, a Schr\"odinger function can
be trivialized and, thereby, its solutions found.  The fact that the
infinitesimal versions of these elementary transformations classically
generate the full canonical algebra is argued to imply that in principle
any canonical transformation can be implemented quantum mechanically.
Issues of operator ordering break the parallel structure
between classical and quantum canonical transformations, so that in general
different transformations are needed to reach the trivial
Schr\"odinger function in each case.  This raises the interesting possibility
of the inequivalence of classical and quantum integrability.

By defining quantum canonical transformations algebraically in terms of a
topological transformation group consisting of ordered expressions in the
quantum variables $q$ and $p$, consistent with the canonical commutation
relations, it was possible to work outside of a specific Hilbert space.
This allowed the use of transformations that are non-unitary when
represented on a particular Hilbert space, either because they change the
measure in the inner product or because they do not define an isomorphism.
While not of importance for evolution, such transformations are important
tools for constructing the solutions to the wave equation. Raising
and lowering operators\cite{InH}, intertwining operators\cite{And3,And4}
and differential realizations of Lie algebras\cite{lam} provide
well-known examples of undeniable importance. As a
by-product of the fact that the quantum canonical
transformations are defined outside of the Hilbert space, they enable the
construction of the general solution of the wave equation, including the
non-normalizable solutions.  This may be important in contexts outside of
quantum mechanics where normalizability of solutions is less important.

This approach also gives new tools for proving the physical equivalence
of quantum theories.  Rather than attempting to factor order a classical
canonical transformation between two theories, one can construct the
quantum transformation in a more systematic fashion.  This enables one to
construct quantum transformations that are non-polynomial factor
orderings of the classical transformation.  A non-trivial example
illustrating this
is the proof of the equivalence of the Moncrief and Witten/Carlip
formulations of 2+1-quantum gravity on the torus\cite{And5}.

Further work is necessary to elaborate the collection of tools for using
the quantum canonical transformations, especially in the many-variable
context.  A fruitful direction for future work is the extension of this
approach to field theory where it may perhaps shed new light on
integrable systems or renormalizability.

\vskip 1.5cm

I would like to thank C.J. Isham for critical comments.
This work was supported in part by a grant from the Natural Sciences and
Engineering Research Council of Canada and Les Fonds FCAR du Qu\'ebec, in part
by the Science and Engineering Research Council of the United Kingdom,
and in part by the National Science Foundation of the United States
under Grant No. PHY89-04035.

\begin{figure}
$$\begin{array}{cc}
\bega{ccc}
p &\mapsto& -q \\
q &\mapsto & p
\enda
&
\sp \check{I}\psi(q)={1\over (2\pi)^{1/2}}\int_{-\infty}^{\infty} dq'
e^{iq q'}\psi(q') \\
&\\[.8cm]
\bega{ccc}
p &\mapsto& q \\
q &\mapsto& -p
\enda
&\sp (I^{-1})\check{\vb}\psi(q)={1\over
(2\pi)^{1/2}}\int_{-\infty}^{\infty} dq'
e^{-iq q'}\psi(q') \\
&\\[.8cm]
\bega{ccc}
p &\mapsto& p -i f(q)\c{q} \\
q &\mapsto& q
\enda
&
\sp (e^{-f(q)})\check{\vb}\psi(q) \\
&\\[.8cm]
\bega{ccc}
p &\mapsto& p \\
q &\mapsto& q +ig(p)\c{p}
\enda
&\sp (e^{-g(p) })\check{\vb}\psi(q) \\
&\\[.8cm]
\bega{ccc}
p &\mapsto& {{\textstyle 1} \over { \textstyle f(q) }
 \c{{\scriptstyle q} }} p \\[.4cm]
q &\mapsto& f(q)
\enda
&
\sp \check{P}_{f(q)}\psi(q)=\psi^{(0)}(f(q)) \\
&\\[.8cm]
\bega{ccc}
p &\mapsto& g(p) \\[.2cm]
q &\mapsto& {{\textstyle 1}\over  {\textstyle g(p) }
\c{{\scriptstyle p} }} q
\enda
&\sp \check{P}_{g(p)}\psi(q)={1\over 2\pi}\int_{-\infty}^\infty d\bq
e^{i\bq q}
\int_{-\infty}^{\infty} dq' e^{-ig(\bq)q'} \psi(q') \\

\end{array}
$$
\caption{Elementary and composite elementary canonical transformations}
\end{figure}


\newpage

\begin{thebibliography}{99}

\bibitem{Dir} P. A. M. Dirac, {\it The Principles of Quantum Mechanics,
4th. ed.}, (Oxford Univ. Press: Oxford, 1958).

\bibitem{Wey} H. Weyl, {\it The Theory of Groups and Quantum Mechanics,
2nd. ed.}, (Dover: New York, 1950, 1931).

\bibitem{lin} C. Itzykson, Comm. Math. Phys. {\bf 4}, 92 (1967);
V. Bargmann, in {\it Analytic Methods in Mathematical
Physics}, edited by R. P. Gilbert and R. G. Newton, (Gordon and Breach: New
York, 1970), p. 27l; M. Moshinsky and C.\ Quesne, J. Math. Phys. {\bf 12} 772
(1971).

\bibitem{Mos} P.A. Mello and M. Moshinsky, J. Math. Phys. {\bf 16} 2017 (1975);
M. Moshinsky, {Groups in Physics} (Les Presses de L'Universit\'e de
Montr\'eal: Montr\'eal, 1979) p. 48; J. Deenen, M. Moshinsky, and T.H.
Seligman, Ann. Phys. (NY) {\bf 127}, 458 (1980), and references therein.

\bibitem{LeS}  F. Leyvraz and T.H. Seligman. J. Math. Phys. {\bf 30},
2512 (1989).

\bibitem{Dee}  J. Deenen, J. Phys. A{\bf 24}, 3851 (1991).

\bibitem{BHJ} M. Born, W. Heisenberg and P. Jordan, Ztschr. f. Phys. {\bf
35}, S. 557 (1926); W. Heisenberg, Math. Ann. {\bf 95}, 683 (1926);
P.A.M. Dirac, Proc. Roy. Soc. {\bf A110}, 561 (1926).

\bibitem{And1} A. Anderson, preprint Imperial-TP-92-93-20/hep-th-9302062
(1993).

\bibitem{And2} A. Anderson, preprint Imperial-TP-92-93-19/hep-th-9302061
(1993).

\bibitem{pseudo} J.J. Kohn and L. Nirenberg, Comm. Pure and Appl. Math.
{\bf 18}, 269 (1965) ; L. H\"ormander, Comm. Pure and Appl. Math. {\bf 18},
501 (1965).

\bibitem{InH} L. Infeld and T.E. Hull, Rev. Mod. Phys. {\bf 23}, 21
(1951).

\bibitem{susy} C.V. Sukumar, J. Phys. A {\bf 18}, L57 (1985); {\bf
18}, 2917 (1985).

\bibitem{And3} A. Anderson, Phys.\ Rev.\ {\bf D37}, 536 (1988); A. Anderson
and R. Camporesi, Comm.\ Math.\ Phys.\ {\bf 130}, 61 (1990).

\bibitem{And4} A. Anderson, Phys.\ Rev.\ {\bf A43}, 4602 (1991).

\bibitem{lam}  See, for example, W. Miller, Jr.,
{\it Lie Theory and Special Functions}, (Academic Press: NY, 1968).;
 N. Ja. Vilenkin, {\it Special Functions and the Theory
of Group Representations}, (Am. Math. Soc.: Providence, Rhode Island,
1968).

\bibitem{Lan} For a discussion of extended phase space in classical
mechanics, see C. Lanczos, {\it The Variational Principles of
Mechanics, 4th ed.} (Dover, New York, 1970).

\bibitem{Rud}  W. Rudin, {\it Functional Analysis}, (McGraw-Hill: New
York, 1973).

\bibitem{Dir2} P.A.M. Dirac, Proc. Camb. Phil. Soc. {\bf 23}, 412 (1926).

\bibitem{Mil} J. Milnor, in {\it Relativity, Groups and Topology, II},
edited by R. Stora and B. DeWitt,
(North Holland: Amsterdam, 1984), 1007.

\bibitem{MoZ} D. Montgomery and L. Zippin, {\it Topological
Transformation Groups}, (Wiley: New York, 1955).

\bibitem{ring} See, for example, T.Y. Lam, {\it A First Course in
Noncommutative Rings}, (Springer-Verlag: New York, 1991).

\bibitem{DuS} N. Dunford and J.T. Schwartz, {\it Linear Operators, Part 1},
(Wiley: New York, 1963).

\bibitem{Gol} H.B. Goldstein, {\it Classical Mechanics, 2nd. ed.}
(Addison-Wesley: Reading, MA, 1980).

\bibitem{vaH} H.J. Groenewold, Physica {\bf 12}, 405 (1946); L. Van Hove,
Acad. roy. Belg. Bull. Classe Sci. M\'em. (5) {\bf 37}, 610 (1951); U.
Uhlhorn, Arkiv. Fysik. {\bf 11}, 87 (1956).

\bibitem{qo}  See, for example, H.-Y. Fan and J. Vanderlinde, Phys. Rev.
{\bf A39}, 2987 (1989); D. Han, Y.S. Kim and M.E. Noz, Phys. Rev. {\bf
A40}, 902 (1989); A. Luis and L.L. Sanchez-Soto, J. Phys. A{\bf 24}, 2083
(1991).

\bibitem{Bro} L.S. Brown, Phys.\ Rev.\ Lett.\ {\bf 66}, 527 (1991).

\bibitem{Ince} E.L. Ince, {\it Ordinary Differential Equations}, (Dover:
New York, 1956, 1926), p. 122.

\bibitem{And5} A. Anderson, Phys.\ Rev.\ {\bf D47}, xxx (1993).

\end{thebibliography}
\end{document}